\documentclass[aps,twocolumn,showpacs]{revtex4}
\usepackage{graphicx}
\usepackage{amsfonts}
\usepackage{amssymb}
\usepackage{amsbsy}
\usepackage{amsmath}
\usepackage{mathrsfs}
\usepackage{latexsym}
\usepackage{natbib}
\usepackage{bm}
\usepackage{color}
\usepackage{braket}
\usepackage{slashed}
\usepackage[normalem]{ulem}

\def\x{\mathbf{x}}

\def\k{\mathrm{k}}

\def\muD{{\bf\mu_D}}

\def\N{\mathcal{N}}

\def\mr{\mathcal{M}}

\begin{document}

\author{Golam Mortuza Hossain}
\email{ghossain@iiserkol.ac.in}

\author{Susobhan Mandal}
\email{sm17rs045@iiserkol.ac.in}

\affiliation{ Department of Physical Sciences, 
Indian Institute of Science Education and Research Kolkata,
Mohanpur - 741 246, WB, India }
 

\date{\today}

\title{Origin of primeval seed magnetism in rotating astrophysical bodies}

\begin{abstract}
We show that a primeval seed magnetic field arises due to spin-degeneracy 
breaking of fermions caused by the dragging of inertial frames in the curved 
spacetime of rotating astrophysical bodies. This seed magnetic field would arise 
even due to electrically neutral fermions such as neutrons. As an example, we 
show that an ideal neutron star rotating at $500$ revolutions per second, 
having mass $0.83$ M$_{\odot}$ and described by an ensemble of degenerate 
neutrons, would have $0.12$ Gauss seed magnetic field at its center arising 
through the breaking of spin-degeneracy. 
\end{abstract}

\maketitle

\section{Introduction}

%
Magnetic fields are observed in the universe on widely different scales. Its 
field strengths are seen to vary from being as small as $10^{-16}$ Gauss in the 
voids of intergalactic medium \cite{neronov2010, essey2011determination, 
dolag2010lower, takahashi2013lower}, around $10^{-6}$ Gauss within a galaxy and 
in the range of $10^{8}- 10^{15}$ Gauss in a rotating neutron star 
\cite{tiengo2013variable}. Given a small seed magnetic field, there 
exists mechanism such as the turbulent dynamo \cite{zhang2009three, 
bucciantini2013fully} that can rapidly amplify the magnetic field strength in 
astrophysical systems. However, a seemingly innocuous yet profound question 
remains poorly understood --- what is the origin of the seed magnetism itself? 
At present there exist largely two different types of model for the seed 
magnetism where the seed field is viewed either as a cosmic relic of the early 
universe physics or as being generated by ionized plasma through the 
astrophysical processes \cite{cho2014origin, science.1120690, 
doi2011generation}. However, these models come with their own set of 
shortcomings such as present field strength being incompatible with early 
universe physics or having insufficient sustenance and coherence of the seed 
field (see \cite{subramanian2019primordial,brandenburg2005astrophysical} for 
an excellent review).

In this article, we show that a natural answer to the question of seed 
magnetism 
arises from two fundamental theories of nature, namely Einstein's general 
theory of relativity and Dirac's theory of fermions when they are put together. 
This answer does not require any new exotic physics but a proper reconciliation 
of the Dirac theory together with the general relativity through the methods of 
quantum field theory in the curved spacetime. In particular, we show that the 
genesis of seed magnetism is a direct consequence of spin-degeneracy breaking of 
fermions caused by the curved spacetime of rotating astrophysical bodies, 
principally due to the \emph{dragging} of inertial frames. This seed magnetism 
would arise even due to electrically neutral fermions such as neutrons.

\section{Fermions in curved spacetime}

%
In Fock-Weyl formulation, dynamics of a free Dirac fermion $\psi$ in a 
generally curved spacetime is described by an invariant action 
\begin{equation}\label{FermionActionI}
S_{\psi} = -\int d^{4}x \sqrt{-g} ~ \bar{\psi} [i \gamma^{a} {e^{\mu}}_a 
\mathcal{D}_{\mu} + m]\psi  ~,
\end{equation}
where Dirac adjoint $\bar{\psi} = \psi^{\dagger}\gamma^0$ and $m$ is the mass of 
the fermion. Here ${e^{\mu}}_a$ are the \emph{tetrad} components defined in 
terms of the global metric $g_{\mu\nu}$ as $g_{\mu\nu} {e^{\mu}}_a {e^{\nu}}_b = 
\eta_{ab}$ where $\eta_{ab} = diag(-1,1,1,1)$ is the Minkowski metric. 
Spin-covariant derivative for the fermion field $\psi$ is defined as 
$\mathcal{D}_{\mu}\psi \equiv \partial_{\mu}\psi + \Gamma_{\mu}\psi $ together 
with the spin connection 
\begin{equation}\label{OmegaMuabDef}
\Gamma_{\mu}  =  -\tfrac{1}{8} \eta_{ac} {e_{\nu}}^c 
(\partial_{\mu} {e^{\nu}}_b + \Gamma^{\nu}_{\mu\sigma} {e^{\sigma}}_b )
~[\gamma^{a}, \gamma^{b}]  ~,
\end{equation}
where $\Gamma^{\nu}_{\mu\beta}$ are the Christoffel connections and $\gamma^{a}$ 
are the Dirac matrices in Minkowski spacetime, satisfying the Clifford algebra 
$\{\gamma^{a},\gamma^{b}\} = - 2\eta^{ab} \mathbb{I}$. The minus sign in front 
of the metric $\eta^{ab}$ here ensures that usual relations $(\gamma^0)^2 = 
\mathbb{I}$ and $(\gamma^k)^2 = -\mathbb{I}$ for $k=1,2,3$ holds true for the 
chosen metric signature. The Lagrangian density $\mathcal{L}$ corresponding to 
the action (\ref{FermionActionI}) is expressed as $S_{\psi} = \int d^{4}x 
\sqrt{-g}~\mathcal{L}~$.

\section{Axially symmetric stationary spacetime}

%
In order to describe the spacetime around a rotating astrophysical body, we 
consider its matter distribution to be axially symmetric and stationary. In 
\emph{natural units} $c = \hbar =1$, the metric due to an axially symmetric, 
stationary and slowly rotating matter distribution can be expressed as 
\cite{hartle1967slowly}
\begin{equation}\label{ASMetric}
ds^2 = - H^2 dt^2 + Q^2 dr^2 + 
r^2 K^2[d\theta^2 +  \sin^2\theta (d\varphi- L dt)^2] ~,
\end{equation}
where metric functions $H$, $Q$, $K$ and $L$ depend only on the coordinates $r$ 
and $\theta$. Here we require these functions to be such that the metric 
(\ref{ASMetric}) is asymptotically flat. The function $L$ represents acquired 
angular velocity of a freely-falling observer from infinity, a phenomena known 
as the \emph{dragging} of inertial frames. For most rotating astrophysical 
bodies the frame-dragging angular velocity is small such that $(r L) \ll 1$. So 
for simplicity we shall ignore terms which are $\mathcal{O}(L^2)$. These metric 
functions are governed by Einstein's equation $G_{\mu\nu} = 8\pi G T_{\mu\nu}$. 
We represent the matter distribution here by a perfect fluid with the 
stress-energy tensor
\begin{equation}\label{StressEnergyTensorPerfectFluid}
T_{\mu\nu} = (\rho + P) u_{\mu} u_{\nu} + P g_{\mu\nu} ~,
\end{equation} 
where $u^{\mu}$ is the 4-velocity of the fluid, $P$ is its pressure and
$\rho$ is its energy density.

\section{Reduced action}

%
In the Einstein equation, the pressure and the energy density are considered to 
be varying functions of the coordinates in general. However, in equilibrium
statistical mechanics, these thermodynamical quantities are considered to be 
uniform within a given system. These two seemingly disparate aspects therefore 
need to be combined consistently by using the notion of \emph{local} 
thermodynamical equilibrium. In essence, one has to consider two distinct 
scales --- the \emph{astrophysical scale} over which Einstein's equation 
operates and the \emph{scale of microscopic physics} over which local 
thermodynamical equilibrium is achieved. In order to ensure a local 
thermodynamical equilibrium, here we consider a small region around every 
spatial point such that the variation of the metric functions within the region 
can be neglected. The small region nevertheless, must contain sufficiently large 
number of degrees of freedom to ensure a thermal equilibrium. The later 
condition, in the context of quantum statistical physics, requires wave 
functions representing those degrees of freedom should be sufficiently 
localized.

For definiteness, let us consider a small spatial box whose center is located 
at the coordinates $(r_{0},\theta_0)$. By defining a new set of local 
coordinates as
$X = Q_0 r \sin \bar{\theta} \cos\bar{\varphi}$, 
$Y = Q_0 r \sin \bar{\theta} \sin\bar{\varphi}$, and
$Z = Q_0 r \cos \bar{\theta}$ where  
$\bar{\theta} = \xi_1 \theta$, $\bar{\varphi} = \xi_0 \varphi$ with 
$\xi_1 = (K_0/Q_0)$, $\xi_0 = ( \xi_1 \sin\theta_0 )/\sin( \xi_1 \theta_0)$, 
the metric (\ref{ASMetric}) can be reduced within the box to be
\begin{equation}\label{MetricInBox}
g_{\mu\nu} = \begin{bmatrix}
- \N^2 & \omega Y & - \omega X & 0\\
\omega Y & 1 & 0 & 0\\
-\omega X & 0 & 1 & 0\\
0 & 0 & 0 & 1
\end{bmatrix}  ~,
\end{equation}
where $\omega = \xi_0 L_0$ and $\N^2 = H^2_0$. Here we have assumed that the 
metric functions take a fixed set of values $H_0$, $Q_0$, $K_0$ and $L_0$ within 
the box such that $Q_0 \ge K_0$ \emph{i.e.} $\xi_1 \le 1$. Additionally, we 
have approximated $\sin^2\theta d\varphi = \sin^2(\theta_0 +\delta\theta) 
d\varphi \approx \sin^2\theta_0 d\varphi$ for all points within the box. 

The non-vanishing tetrad components corresponding to the metric 
(\ref{MetricInBox}), can be expressed as ${e^X}_1 = {e^Y}_2 = {e^Z}_3 = 1$, 
${e^t}_0 = \mathcal{N}^{-1}$, ${e^X}_0 = -\omega Y \mathcal{N}^{-1}$, ${e^Y}_0 = 
 \omega X \mathcal{N}^{-1}$ whereas non-vanishing components of the inverse 
tetrad ${e_{\mu}}^a$ can be written as ${e_X}^1 = {e_Y}^2 = {e_Z}^3 = 1$, 
${e_t}^0 = \mathcal{N}$, ${e_t}^1 = \omega Y$, ${e_t}^2 = -\omega X$.
Non-vanishing components of the Christoffel connection are given by 
$\Gamma^X_{tY} = \Gamma^X_{Yt} = -\Gamma^Y_{tX} = -\Gamma^Y_{Xt} = \omega$. 
Consequently, only non-vanishing component of the spin-connection $\Gamma_{\mu}$ 
is
\begin{equation}\label{SpinConnectionInBox}
 \Gamma_{t} = - \frac{\omega}{4} [\gamma^{1},\gamma^{2}] 
 = \frac{i \omega}{2}\Sigma_3  ~,
\end{equation}
where $\Sigma_3 = \sigma^3 \otimes\mathbb{I}_2$ with $\sigma^3$ being the third 
Pauli matrix. Therefore, the Dirac action (\ref{FermionActionI}) within the box 
reduces to
\begin{equation}\label{FermionActionInBox}
S_{\psi} = -\int d^{4}x ~\bar{\psi} [ i \gamma^0 \partial_t + 
\mathcal{N} (i\gamma^k \partial_k + m) - \omega \gamma^0 \hat{J}_Z ]\psi  ~,
\end{equation}
where $\hat{J}_Z = (\hat{L}_Z + \tfrac{1}{2} \Sigma_3)$ with $\hat{L}_Z = 
-i(X\partial_{Y} - Y\partial_{X})$. We note that $\hat{J}_Z$ can be naturally 
interpreted as the total angular momentum operator associated with the dragging 
of inertial frames where $\hat{L}_Z$ is the orbital angular momentum operator 
and $\sigma^3$ is the third Pauli matrix which is the spin operator along 
$Z$-direction. One may arrive at the reduced action (\ref{FermionActionInBox}) 
also by considering the transformation of the spinor field  $\psi(x) \mapsto 
e^{i\hat{J}_{Z}\omega t}\psi(x)$ under a rotation around $Z$ axis with angular 
velocity $\omega$, starting from a non-rotating configuration.

\section{Partition function}

%
We consider an ensemble of fermions in the box which is in a local 
thermodynamical equilibrium. Here we define the scale of temperature $T$ in the 
frame of an asymptotic observer where $\mathcal{N}\to 1$ and $\omega\to 0$. 
This choice allows us to treat the reduced action (\ref{FermionActionInBox}) as 
an \emph{effective} action written in the Minkowski spacetime. It leads to 
a simpler computation as well as it helps to avoid the issues related to Wick 
rotation in a general curved spacetime \cite{visser2017wick}. The effective 
action nevertheless includes the effects of curved spacetime through the fixed 
parameters $\mathcal{N}$ and $\omega$ in the given box. In order to compute the 
partition function, here we follow similar methods as used by the authors for 
computing equation of states in the curved spacetime of spherical stars 
\cite{hossain2021equation, hossain2021higher}. Using coherent states of the 
Grassmann fields \cite{laine2016basics, das1997finite, kapusta1989finite}, the 
partition function corresponding to the action (\ref{FermionActionInBox}) can be 
expressed as 
$\mathcal{Z}_{\psi} = \int\mathcal{D} \bar{\psi} \mathcal{D}\psi \ 
e^{-S_{\psi}^{\beta}}$ where $S_{\psi}^{\beta} = \int_{0}^{\beta}d\tau\int 
d^{3}x (\mathcal{L}^{E} - \mu\bar{\psi}\gamma^{0}\psi)$. Here $\mu$ is the 
chemical potential of the fermion and $\beta = 1/k_{B}T$ with $k_{B}$ being the 
Boltzmann constant. The Euclidean Lagrangian density is obtained through a Wick 
rotation as $\mathcal{L}^{E} = -\mathcal{L}(t\rightarrow -i\tau)$.

In a thermal equilibrium, the fermion field $\psi$ is subjected to the 
\emph{anti-periodic} boundary condition $\psi(\tau,\x) = -\psi(\tau+\beta,\x)$. 
Consequently, in Fourier domain the Dirac field $\psi$ can be written as
\begin{equation}\label{FermionicFourier}
\psi(\tau,\x) = \frac{1}{\sqrt{V}} \sum_{l,\k} ~e^{-i(\omega_l\tau + 
\k\cdot\x)} \tilde{\psi}(l,\k)  ~,
\end{equation}
where Matsubara frequencies are $\omega_l = (2l+1) \pi ~\beta^{-1}$ with $l$ 
being an integer and $V = \int d^3x\sqrt{-\eta}$ is the volume of the box. 
Using the reduced action (\ref{FermionActionInBox}) it is convenient to express 
the partition function as $\ln\mathcal{Z}_{\psi} = \ln\mathcal{Z}_0 + 
\ln\mathcal{Z}_L$ where $\mathcal{Z}_0 = \int\mathcal{D}\bar{\psi} 
\mathcal{D}\psi ~e^{-S^{\beta}_0}$ and
\begin{equation}\label{EuAction0II}
S^{\beta}_0 = \sum_{l,\k} ~\bar{\tilde{\psi}}~\beta
\left[ \slashed{p} + \bar{m} \right]  \tilde{\psi} ~,
\end{equation}
with $\bar{m} =  m \mathcal{N}$,  $\slashed{p} = \gamma^{0}(i\omega_l - \mu - 
\frac{\omega}{2} \Sigma_3) + \gamma^{k} (\k_k \mathcal{N})$. On the other hand, 
$\mathcal{Z}_L$ can be expressed as a perturbative series $\ln\mathcal{Z}_L = 
\ln ( 1 + \sum_{l=1}^{\infty} \frac{\omega^l}{l!} \langle (-S^{\beta}_L)^l 
\rangle)$ where $S_{L} =  \int_{0}^{\beta}d\tau\int d^{3}x \bar{\psi}[\gamma^{0} 
\hat{L}_{Z}]\psi$. It can be shown that $\ln\mathcal{Z}_L \sim 
\mathcal{O}(\omega^2)$ and hence it can be ignored. By employing the Dirac 
representation of $\gamma^a$ matrices and the results of Gaussian integral over 
Grassmann fields, one can evaluate the partition function as $\ln\mathcal{Z}_{0} 
= \ln\mathcal{Z}_{+} + \ln\mathcal{Z}_{-}$ where 
\begin{equation}\label{LogPartitionFunctionPM2}
\ln\mathcal{Z}_{\pm} = \sum_{\k} \left[ 
\ln\big(1 + e^{-\beta(\varepsilon - \mu_{\pm})} \big) 
+ \ln\big(1 + e^{-\beta(\varepsilon + \mu_{\pm})} \big)
\right] ~,
\end{equation}
with $\varepsilon^2 = \N (\k^2 + m^2)$ and $\mu_{\pm} = \mu \pm (\omega/2)$. To 
arrive at the expression (\ref{LogPartitionFunctionPM2}), formally divergent 
terms including the zero-point energy are dropped. In the equation 
(\ref{LogPartitionFunctionPM2}), the first and the second terms correspond to 
the \emph{particle} and  \emph{anti-particle} sectors respectively. Here we 
consider the rotating astrophysical body to be made of only \emph{particles} for 
which the partition function becomes
\begin{equation}\label{LogPartitionFunctionFinal}
\ln\mathcal{Z}_{\psi} = \sum_{\k} \left[ 
\ln\big(1 + e^{-\beta(\varepsilon - \mu_{+})} \big) +
\ln\big(1 + e^{-\beta(\varepsilon - \mu_{-})} \big) \right] ~.
\end{equation}
The presence of Pauli matrix $\sigma^3$ in $\slashed{p}$ leads the partition 
function (\ref{LogPartitionFunctionFinal}) to split up in two parts with 
different energy levels corresponding to the spin-up and the spin-down fermions 
respectively which in turns breaks the spin-degeneracy of fermions. In the 
absence of dragging of inertial frames \emph{i.e.} if one takes $\omega\to 
0$ limit, then the partition function (\ref{LogPartitionFunctionFinal}) reduces 
to its usual form with the spin-degeneracy factor of $2$. However, for any 
non-vanishing $\omega$, there exists a gap between the energy levels of the 
spin-up and the spin-down fermions having same $\varepsilon$. The energy gap 
nevertheless is very small. For example, if the frame-dragging angular velocity 
$\omega$ is one revolution per second then for neutrons the ratio $(\mu_{+} - 
\mu_{-})/\mu = (\omega/\mu) < (\omega/m) \sim 10^{-24}$.

\section{Primeval magnetic field}

%
The number density of fermions that follows from the partition function 
(\ref{LogPartitionFunctionFinal}) as $n = (\beta V)^{-1} (\partial
\ln\mathcal{Z}_{\psi}/\partial\mu)$, can be expressed as
\begin{equation}\label{NumberDensityPM}
n = n_{+} + n_{-} ~,~~  \text{with} ~~ n_{\pm} = 
\frac{1}{V} \sum_{\k} \frac{1}{e^{\beta(\varepsilon-\mu_{\pm})} + 1} ~. 
\end{equation}
We note that the number densities for the spin-up and the spin-down fermions 
are different. Consequently, it gives rise to a net magnetic moment 
$\mathfrak{M} = \muD (n_{+} - n_{-})$ where $\muD$ is the \emph{magnitude} of 
the magnetic moment of a spin-up Dirac fermion. The corresponding magnetic 
moment then can be expressed as
\begin{equation}\label{MagneticMoment}
\mathfrak{M} = \muD \frac{\beta\omega}{V} \sum_{\k} 
\frac{e^{\beta(\varepsilon-\mu)}}{(e^{\beta(\varepsilon-\mu)} + 1)^2} 
+ \mathcal{O}(\omega^2) ~.
\end{equation}
The magnetic field arising due to the magnetic moment (\ref{MagneticMoment}), 
can be obtained by computing the magnetic \emph{susceptibility} of the 
fermionic matter. The susceptibility represents the response of spin degrees of 
freedom in orienting themselves along the direction of an external magnetic 
field $B$ and is defined as $\chi = (\partial\mathfrak{M}/\partial B)_{|B=0}$. 
In order to compute the magnetic susceptibility, here we consider a test 
magnetic field $B$ along the $Z$-direction. The coupling between an 
electrically \emph{neutral} fermion field $\psi$ and the electromagnetic field 
$A_{\mu}$ is described by Pauli-Dirac interaction term $S_I = \int d^4x 
\sqrt{-g} ~\bar{\psi}[ \frac{1}{2} \muD \sigma^{\mu\nu} F_{\mu\nu}\Big] \psi$ 
where $\sigma^{\mu\nu} =  \frac{i}{2} {e^{\mu}}_a {e^{\nu}}_b [\gamma^a, 
\gamma^b]$ and $F_{\mu\nu} = \partial_{\mu} A_{\nu} - \partial_{\nu}A_{\mu}$. 
With a gauge choice $A_{\mu} = (0,0,B X, 0)$, the interaction term reduces 
within the box as
\begin{equation}\label{ReducedActionSpinB}
S_{I} = \int d^{4}x \bar{\psi}\Big[\muD \N B ~ \Sigma_3 \Big] \psi ~.
\end{equation} 
For the \emph{particle} sector, the contribution from (\ref{ReducedActionSpinB}) 
effectively alters $\omega \to \tilde{\omega} = \omega + 2\muD \N B$ in the 
action (\ref{FermionActionInBox}) whereas for the \emph{anti-particle} sector it 
changes $\omega \to \tilde{\omega} = \omega - 2\muD \N B$. In the context of 
zero-temperature field theory, it would directly imply that the effect of 
dragging of inertial frames on the particle sector can be traded off by an 
external magnetic $B = \omega/(2\muD \N)$. This aspect holds true even for 
thermal field theory with non-zero temperature in the leading 
order in $\omega$. The magnetic susceptibility now can be computed as $\chi = 
2\muD \N (\partial\mathfrak{M}/\partial \omega)$ leading to
\begin{equation}\label{MagneticSusceptibility}
\chi = 2 \muD^2 \frac{\beta \N}{V} \sum_{\k} 
\frac{e^{\beta(\varepsilon-\mu)}}{(e^{\beta(\varepsilon-\mu)} + 1)^2} 
+ \mathcal{O}(\omega) ~.
\end{equation}
Therefore, the resultant magnetic field in the small box is given by
\begin{equation}\label{MagneticField}
B = \frac{\mathfrak{M}}{\chi} = \frac{1}{2\muD} \frac{\xi_0 L_0}{H_0} 
+ \mathcal{O}(L_0^2) ~.
\end{equation}
The equation (\ref{MagneticField}) establishes that a primeval \emph{finite} 
magnetic field 
arises spontaneously due to the spin-degeneracy breaking of fermions in the 
curved spacetime of a rotating astrophysical body. This genesis of primeval 
magnetism, led by a non-vanishing frame-dragging angular velocity, works even 
for electrically neutral fermions such as neutrons. The resultant primeval 
magnetic field thus can act like a \emph{seed} magnetic field which can be used 
for subsequent amplification by other astrophysical processes. In a scenario 
with multiple species of fermions one needs to consider their combined 
contributions to the magnetic moment (\ref{MagneticMoment}).

\section{Slowly rotating neutron star}

\begin{figure}
\includegraphics[height = 5.8cm]{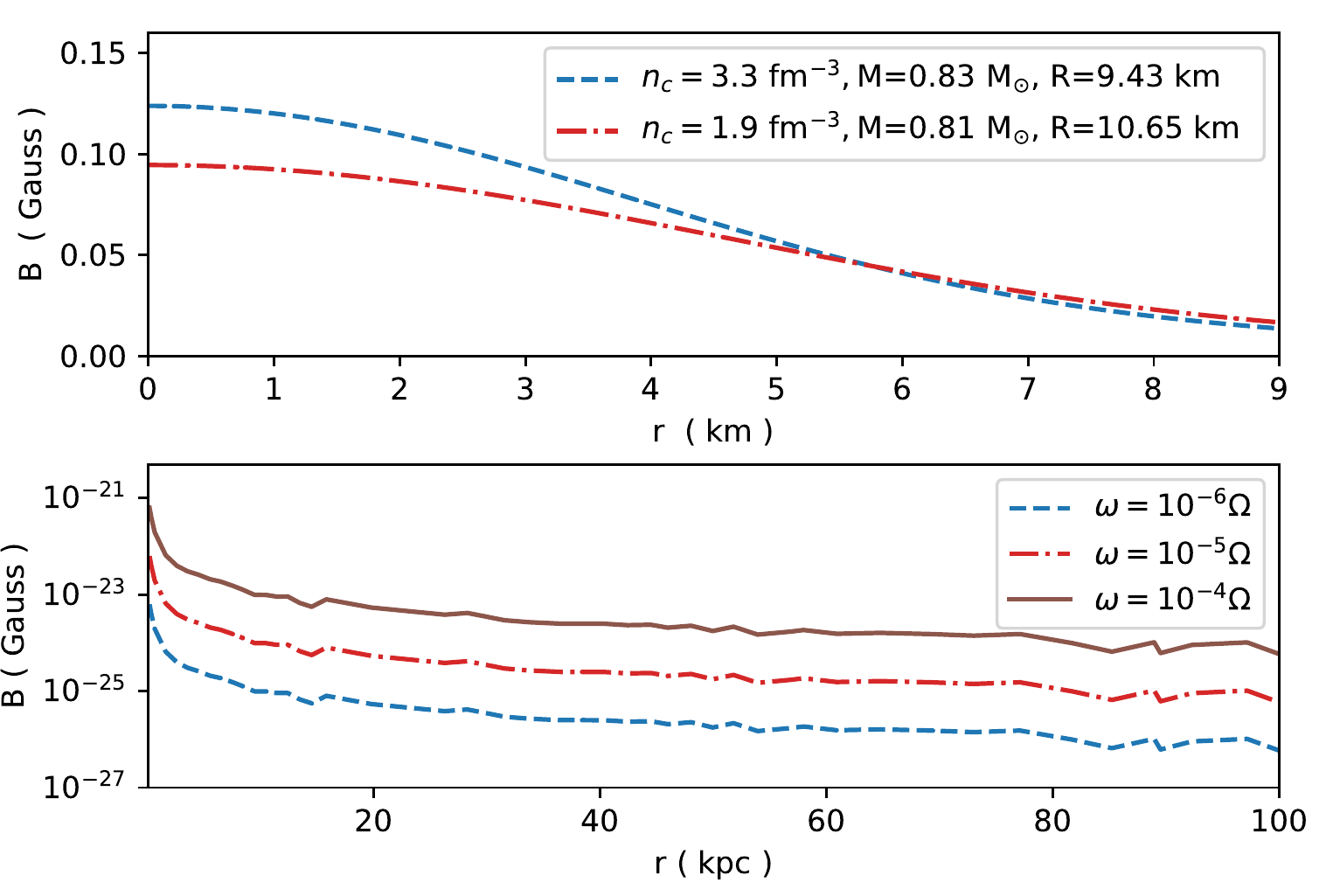}
\caption{The upper panel shows the seed magnetic field inside an ideal neutron 
star whose stellar fluid is revolving 500 times a second. For a central neutron 
number density of $3.3$ fm$^{-3}$, the star has a seed magnetic field of around 
$0.12$ Gauss at its center. Here we have taken $\muD = 9.66\times10^{-31}$ 
J/Gauss for neutrons. The lower panel shows estimated seed magnetic field from 
the observed rotation curve of the Milky Way galaxy at current epoch for 
different compactness.}
\label{fig:ns_galaxy_seed_field}
\end{figure}

%
For quantitative predictions, we now consider a \emph{slowly} rotating ideal 
neutron star (NS) whose degenerate core consists of an ensemble of 
non-interacting neutrons. This choice also highlights the fact that the seed 
magnetism arises even due to electrically neutral fermions. In general, a 
rotating star has the shape of an oblate spheroid. However, for a slowly 
rotating star its mass and radius can 
be decomposed into a `spherical' part and a set of non-spherical perturbative 
corrections which are $\mathcal{O}(\omega^2)$ \cite{hartle1967slowly}. So for 
spherical part we may expand $\xi_0$ as $\xi_0 = 1 + \Delta(\theta_0)$ and 
consider only the leading term. Consequently, the spacetime metric of a slowly 
rotating star can be obtained from the axially symmetric, stationary metric 
(\ref{ASMetric}) with the following choices of the metric functions
\begin{equation}\label{AS2SRSMetric}
 H = e^{\Phi(r)} ~,~  Q = e^{\nu(r)} ~,~ K = 1 ~,~ L = \omega(r)  ~.
\end{equation}
The equations (\ref{MagneticField}, \ref{AS2SRSMetric}) then lead to a seed 
magnetic field $B = \omega e^{-\Phi}/(2\muD)$. We consider a stellar fluid 
which is uniformly rotating with angular velocity $\frac{d\varphi}{dt} = \Omega$ 
with respect to an observer at infinity. Then the 4-velocity of the stellar 
fluid is $u^{\mu} = (e^{-\Phi}~,0,0,\Omega e^{-\Phi})$ and non-vanishing 
components of its co-vector $u_{\mu}$ are $u_t = -e^{\Phi}$ and $u_{\varphi} 
= r^2\sin^2\theta (\Omega-\omega) e^{-\Phi}$. Here slow rotation means $\Omega 
R \ll 1$ with $R$ being radius of the `spherical' part. The metric function 
$\omega$ is governed by $t-\varphi$ component of Einstein's equation given by 
\cite{hartle1967slowly}
\begin{equation}\label{OmegaEqn}
\frac{1}{r^{4}}\frac{d}{dr}\left(r^{4}j\frac{d \omega}{dr}\right) + 
\frac{4}{r}\frac{dj}{dr}(\omega - \Omega) = 0 ~,
\end{equation}
where $j=e^{-(\nu+\Phi)}$. The metric function $\nu$ satisfies $e^{-2\nu} = 1 - 
2 G \mr/r$ with $d\mr = 4\pi r^{2}\rho dr$ whereas the equations for $\Phi$ and 
the pressure $P$ are given by
\begin{equation}\label{TOVEqn}
\frac{d\Phi}{dr} = \frac{G(\mr + 4\pi r^3 P)}{r(r - 2 G \mr)} ~~,~~
\frac{dP}{dr} = - (\rho + P) \frac{d\Phi}{dr} ~.
\end{equation}
The interior metric solutions must match with the exterior vacuum solutions 
$e^{2\Phi} = 1 - 2GM/r$ and $\omega = 2GJ/r^3$ at the star surface. It leads 
to the following boundary conditions $e^{2\Phi(R)} = (1 - 2 G M/R)$ and 
$\omega'(R) = -3\omega(R)/R$ where $\omega' = (d\omega/dr)$, mass $M = \mr(R)$ 
and angular momentum is $J$. The regularity of the equation (\ref{OmegaEqn}) 
additionally demands $\omega'(0) = 0$. By using the curved spacetime of a 
slowly rotating star, the equation of state for an ensemble of degenerate 
neutrons has been computed in an accompanying article \cite{SRSEOS2022}. 
Additionally, a numerical method for solving corresponding Einstein's equation 
together with the constraints is also described there. By using the 
said method, resultant seed field $B$ inside an ideal neutron star is 
plotted in the FIG. \ref{fig:ns_galaxy_seed_field}. 
The presence of a \emph{non-zero} seed magnetic field is crucial in 
a newly born proto-neutron star (PNS) where seed field can be amplified almost 
exponentially by turbulent dynamo mechanism. For example, a seed field of 
$0.1$ Gauss, as seen here, can be amplified to around $10^{11}$ Gauss 
\cite{naso2008magnetic}, a typically observed field strength in neutron stars. 
Further, due to the expected \emph{differential} rotation inside a PNS, the seed 
field itself could be stronger.

\section{Rotating Galaxy}

The seed magnetism studied here would arise in all rotating astrophysical bodies 
including the galaxies. For example, the angular velocity 
$\Omega$ of the Milky Way galaxy varies from $10^{-14}$ to $10^{-17}$ 
rev.s$^{-1}$ in the distance range $0.2 - 100$ kpc from the galactic center, as 
implied by its observed \emph{rotation curve} \cite{bhattacharjee2014rotation}. 
It also implies the dominant presence of \emph{dark matter} in the galaxy where 
stellar mass contributes only around $4\%$ of its total mass of $\sim 10^{12} 
M_{\odot}$ \cite{mcmillan2016mass}. The corresponding frame-dragging velocity 
$\omega$ can be obtained in principle by solving Einstein equation with a 
non-trivial galactic matter distribution. However, a lower bound on $\omega$ can 
be obtained by using the vacuum solution $\omega = 2GJ/R^3$ at the edge of 
galaxy with a spherical dark matter hallo. Assuming Newtonian expression of 
angular momentum $J$, we can approximate $\omega \sim (2GM/R) ~\Omega$. The 
observed mass $M$ of the Milky Way implies a compactness ratio $2GM/R \simeq 
10^{-6}$ at $R=100$ kpc. Inside the galaxy the compactness ratio is expected to 
be higher. For different choices of compactness ratio, the galactic seed 
magnetic field is plotted  in the FIG. \ref{fig:ns_galaxy_seed_field}. As 
earlier, a seed field is important in a proto-galaxy in the early universe. 
Using the covariant scaling of magnetic field as $B(1+z)^2$, the seed field in 
the proto-galaxy at a redshift $z\sim 9$, would be between $10^{-19}-10^{-24}$ 
Gauss. In comparison, the astrophysical battery mechanism of Biermann produces 
$\sim 10^{-21}$ Gauss seed field at a proto-galactic stage which has been shown 
to be sufficient  to produce presently observed microgauss magnetic field 
\cite{kulsrud1997protogalactic}. Inside a galaxy, additional smaller scale seed 
fields of varied magnitude would be created due to subsequent formation of 
different rotating stars.

\section{Discussions}

%
We have shown that the frame-dragging effect \emph{unavoidably} leads 
to a seed magnetic field due to the spin-degeneracy breaking of fermions. The 
seed field is shown to be sufficient in strength and is coherent over a 
very large length scale to be astro-physically relevant. The conservation of 
angular momentum ensures that the seed field is sustained over a very long 
period of time. Further, the seed field arises only after rotating bodies are 
formed during the structure formation, without affecting the early universe 
physics. Therefore, the mechanism shown here are free from the shortcomings of 
other studied mechanisms in the literature.
For simplicity here we have considered an ensemble of non-interacting 
neutrons. An ensemble of interacting neutrons would lead to relatively higher 
mass stars \cite{hossain2021higher} but without significantly affecting the 
result shown here. We note that particle and anti-particle asymmetry for 
neutrinos moving around a rotating black hole has been studied earlier 
\cite{mukhopadhyay2005neutrino}.

\begin{acknowledgments}
SM thanks IISER Kolkata for support through a doctoral fellowship. GMH 
acknowledges support from the grant no. MTR/2021/000209 of SERB, Government of 
India.
\end{acknowledgments}

%

\end{document}